# 3D-printed facet-attached optical elements for beam shaping in optical phased arrays

Stefan Singer,[1,6] Yilin Xu,[1,2] Sebastian Tobias Skacel,[1,3] Yiyang Bao,[1,2] Heiner Zwickel,[1] Pascal Maier,[1,2] Lukas Freter,[1] Philipp-Immanuel Dietrich,[3] Mathias Kaschel,[4] Christoph Menzel,[5] Sebastian Randel,[1] Wolfgang Freude,[1] and Christian Koos[1,2,3,7]

[1]*Institute of Photonics and Quantum Electronics (IPQ), Karlsruhe Institute of Technology (KIT), Engesserstrasse 5, 76131 Karlsruhe, Germany*
[2]*Institute of Microstructure Technology (IMT), Karlsruhe Institute of Technology, Hermann-von-Helmholtz-Platz 1, 76344 Eggenstein-Leopoldshafen, Germany*
[3]*Vanguard Automation GmbH, Gablonzer Strasse 10, 76185 Karlsruhe, Germany*
[4]*Institut für Mikroelektronik Stuttgart (IMS CHIPS), Allmandring 30a, 70569 Stuttgart, Germany*
[5]*SICK AG, Erwin-Sick-Strasse 1, 79183 Waldkirch, Germany*
[6]*stefan.singer@kit.edu*
[7]*christian.koos@kit.edu*

**Abstract:** We demonstrate an optical phased-array (OPA) equipped with a 3D-printed facet-attached element for shaping and deflection of the emitted beam. The beam shaper combines freeform refractive surfaces with total-internal-reflection (TIR) mirrors and is *in-situ* printed to edge-emitting waveguide facets using high-resolution multi-photon lithography, thereby ensuring precise alignment with respect to on-chip waveguide structures. In a proof-of-concept experiment, we achieve a grating-lobe free steering range of $\pm 30°$ and a full-width-half-maximum (FWHM) beam divergence of approximately $2°$. The concept opens an attractive alternative to currently used grating structures and is applicable to a wide range of integration platforms.

## 1. Introduction

Optical phased-arrays (OPA) open a promising path towards compact robust beam scanners that do not contain any mechanically moving parts and that are, e.g., key to advanced light detection and ranging (LiDAR) sensors [1-3]. In general, OPA rely on rather large photonic integrated circuits (PIC) that are usually mounted into densely packed assemblies, thus making edge emission difficult and rendering light emission perpendicular to the chip surface the only practical option [1,4]. This requires efficient coupling of light guided in the planar PIC to a well-defined free-space beam propagating in an out-of-plane direction. In this context, grating structures etched into the in-plane waveguide array have become a mainstay for high index-contrast silicon photonics, which can rely on high lithographic resolution [1-8]. However, such grating couplers require tight process control, especially when well-defined beam emission profiles need to be maintained. Moreover, the efficiency of grating couplers is impaired by the fact that light is diffracted both to the top and to the bottom, unless more complex structures such as bi-layer waveguides [5] are used. If the downward-radiation is redirected to the top, e.g., by a back-side mirror [6], unwanted intensity fringes may appear in the emitted beam along the scanning direction. It should also be noted that emission perpendicular to the chip surface is difficult to achieve by grating structures, since this would unavoidably lead to unwanted coupling of power to the counterpropagating in-plane waveguide mode. Some integration platforms suited for implementation of OPA do not even contain grating structures and hence have to rely on edge emission [9,10].

    In this paper, we demonstrate an alternative approach for beam shaping and for deflecting light from a planar integrated OPA to an out-of-plane direction. The concept relies on 3D-

printed facet-attached elements that combine freeform refractive surfaces with total-internal-reflection (TIR) mirrors. These elements are *in-situ* printed to edge-emitting waveguide facets using high-resolution multi-photon lithography, thereby ensuring highly precise alignment with respect to on-chip PIC structures. In a proof-of-concept experiment, we design, implement, and characterize an edge-emitting silicon photonic (SiP) OPA with a facet-attached beam-shaping structure, offering a scanning range of $\pm 30°$. The full width at half maximum (FWHM) beam divergence is $2.1°$ along and $1.9°$ perpendicular to the scanning direction. The emission efficiency of our current 3D-printed beam-shaping elements amounts to 72 % (1.4 dB loss) and can well compete with best-in-class $Si_3N_4$-based dual-level gratings [11], which occupy considerable on-chip footprint and which rely on rather complex sequences of patterning and deposition steps. In contrast to conventional grating structures, the emission direction of 3D-printed beam-shaping elements can be adapted to any angle, and the concept can be widely applied to practically any integration platform that offers edge-emitting facets.

## 2. Concept and design of 3D-printed beam shapers

Figure 1(a) shows the concept of an integrated OPA-based beam scanner that relies on 3D-printed facet-attached optical elements for beam shaping, see Inset (1). The optical part of the assembly comprises a laser-diode (LD) chip, a photonic integrated circuit (PIC) containing the optical phased-array (OPA), and the 3D-printed optical beam shaper. The beam shaper matches the emitted beam to different application requirements and is printed to the device facet *in-situ* using high-resolution multi-photon lithography. The phases in the branches of the OPA are thermally tuned by on-chip heaters. A printed circuit board (PCB) carries peripheral electronics such as the LD driver, digital-to-analog converters (DAC) that feed the heaters, and microcontrollers (µC) that adjust the phases required for beam steering. The PCB is electrically connected to the PIC by metal wirebonds. The PIC and the LD chip are mounted on a common metallic submount serving as a heat sink and can be efficiently connected by 3D-printed photonic wirebonds (PWB) that eliminate the need for active alignment during the assembly process [12,13].

The tight integration of the PIC into the optoelectronic assembly renders light emission to a surface-normal direction the most practical option. This is accomplished by the 3D-printed beam shaper at the output facet of the OPA, which comprises a cylindrical lens to collimate the emitted light perpendicular to the scanning direction and a total-internal-reflection (TIR) mirror to redirect the beam to a surface-normal direction, see Inset (1) of Fig. 1(a). The implementation of the 3D-printed beam shaper is illustrated in more detail in Fig. 1(b), which shows a side-view of the cylindrical collimation lens and the redirecting mirror. In the following, $\Psi$ and $\Theta$ denote the far-field angles along and perpendicular to the scanning direction, respectively, both measured with respect to the surface normal of the PIC.

To demonstrate the viability of 3D-printed beam shapers for OPA, we implement a proof-of-concept integrated system relying on an OPA with $N = 32$ output channels, fabricated on the silicon photonics (SiP) platform of IMS CHIPS (Stuttgart, Germany) [14]. The layer stack of the photonic chip is illustrated in Fig. 1(b). The silicon-on-insulator (SOI) device layer has a standard thickness of $220\,\text{nm}$ and is optically isolated from the silicon handle wafer by a $3\,\mu\text{m}$ thick buried oxide layer (BOX). The waveguides are overclad by a $2.1\,\mu\text{m}$-thick layer of sputtered silicon dioxide, and the light-emitting facets are created by etching deep trenches into the top oxide. Note that the light-emitting tips of the SiP waveguides (WG) lie entirely within the top oxide, approximately $d_{\text{fac}} \approx 1\,\mu\text{m}$ behind the deep-etched sidewalls of these trenches to safely avoid unwanted uncovering of the waveguide tips during trench formation. For a precise lens design, this spacing has to be taken into account, since the refractive index of the $SiO_2$ layer surrounding the tip ($n = 1.44$ @ $\lambda_0 = 1.55\,\mu\text{m}$) deviates slightly from that of the 3D-printed lens body ($n = 1.53$ @ $\lambda_0 = 1.55\,\mu\text{m}$), leading to slight refraction of light at the interface.

Heaters and contact pads for metal wirebonds are implemented by a AlSiCu metal layer that features a thickness of $100\,\text{nm}$ and that is deposited on top of the cladding. The standard trace width for on-chip electrical connections amounts to $20\,\mu\text{m}$, while the trace width of the heater sections is only $1\,\mu\text{m}$, thus leading to an increased electrical resistance and therefore to heating of the silicon waveguides below.

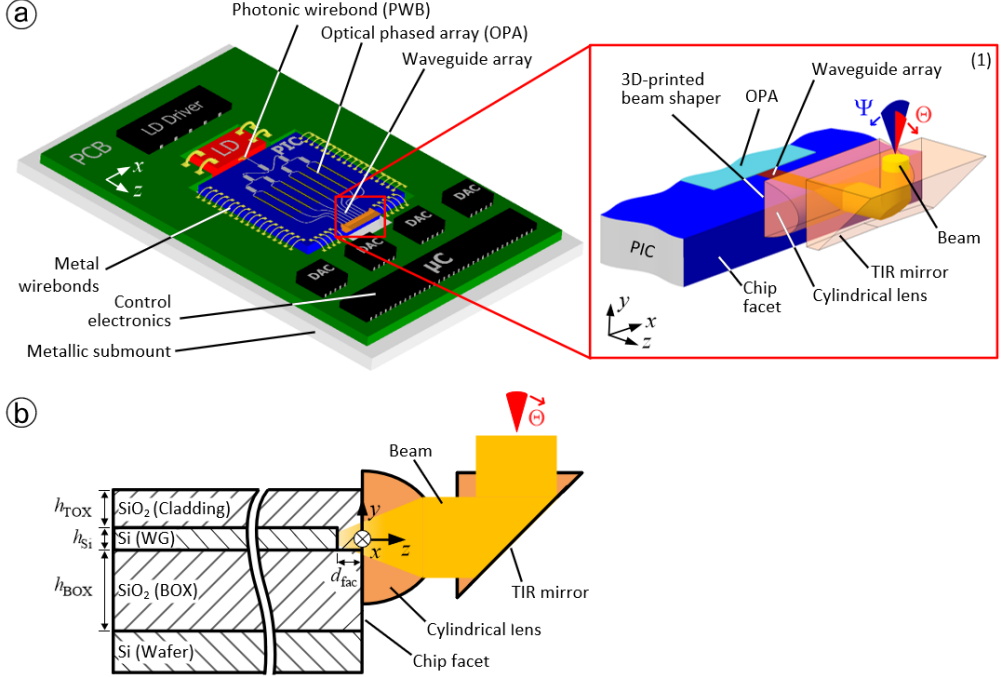

**Fig. 1**. Vision of an integrated beam scanner based on an optical phased-array (OPA) with a 3D-printed beam shaper. **(a)** Artist's impression of a beam scanner system consisting of an OPA on a silicon photonic integrated circuit (PIC), a laser-diode (LD) chip, and control electronics such as LD drivers, digital-to-analog converters (DAC), or microcontroller (µC). All optical components are assembled on a common metallic submount that also functions as a heat sink. Laser and OPA are optically connected via a photonic wirebond (PWB). The OPA waveguide facets emit a beam, which is then shaped by a 3D-printed cylindrical collimation lens and a total-internal reflection (TIR) mirror that redirects the light to the top, see Inset (1) and Subfigure (b). Phase shifters in the OPA waveguides are used to steer the beam in the chip plane ($(x, z)$-plane), which, after the beam shaper, results in a steering along angle $\Psi$. The beam divergence in the direction $\Theta$ perpendicular to the scan direction remains nominally unchanged. Inset (1): Details of the 3D-printed facet-attached beam shaper, comprising the cylindrical collimation lens and the TIR redirecting mirror. **(b)** Cross-section of the PIC stack and the beam shaper. The buried oxide (BOX) has a thickness of $h_{\text{BOX}} = 3\,\mu\text{m}$, and the silicon photonic (SiP) waveguides (WG) are $h_{\text{Si}} = 220\,\text{nm}$ thick. The waveguides are covered with a top oxide layer (thickness $h_{\text{TOX}} = 2.1\,\mu\text{m}$) and end in a distance of $d_{\text{fac}} \approx 1\,\mu\text{m}$ from the chip facet.

To characterize the phase shifters, we use Mach-Zehnder interferometer (MZI) test structures, revealing a thermal time constant of $\tau \approx 6\,\mu\text{s}$ and a heating power of $P_\pi \approx 15\,\text{mW}$ required for a phase shift of $\Delta\varphi = \pi$. To minimize thermal coupling of neighboring phase shifters, trenches are etched into the silicon dioxide cladding and BOX between adjacent thermal phase shifters (not shown in Fig. 1).

Photos and microscope images of the proof-of-concept assembly are shown in Fig. 2. Laser light with a free-space wavelength of $\lambda_0 = 1.55\,\mu\text{m}$ is coupled to the SiP OPA via a waveguide facet at the chip edge using a lensed fiber, see Fig. 2(a). The coupled light is split up evenly into $N = 32$ waveguides using a tree of $2 \times 2$ multi-mode interference (MMI) couplers, see Fig. 2(b). A subset of the 32 waveguides can be fed via grating couplers (GC) within the MMI splitting tree. The phases of the propagating waves in all 32 waveguides can be tuned individually by more than $\Delta \varphi = 2\pi$ using the aforementioned thermo-optic phase shifters, see Fig. 2(c). After the optical phase shifter sections, the waveguides are brought to a pitch of $d = 1.5\,\mu m$, which is slightly smaller than the optical operating wavelength $\lambda_0$, see Figs. 2(d) and 2(e). This leads to a steering range without secondary main lobes (grating lobes) of $\Psi \approx -30°...30°$ and a beam with a theoretical full width at half maximum (FWHM) divergence in the steering direction of $\Delta \Psi_{FWHM} \approx 1.64°$ when steered to $\Psi = 0°$ [15]. Note that the steering range could be further increased by using sub-wavelength spacing of the edge-emitting waveguide facets as demonstrated in [16,17]. Note also that the chip footprint shown in Fig. 2(a) can be significantly reduced as most of the chip area is occupied by partially unnecessary electrical routing, designed for simplified packaging via metal wirebonds.

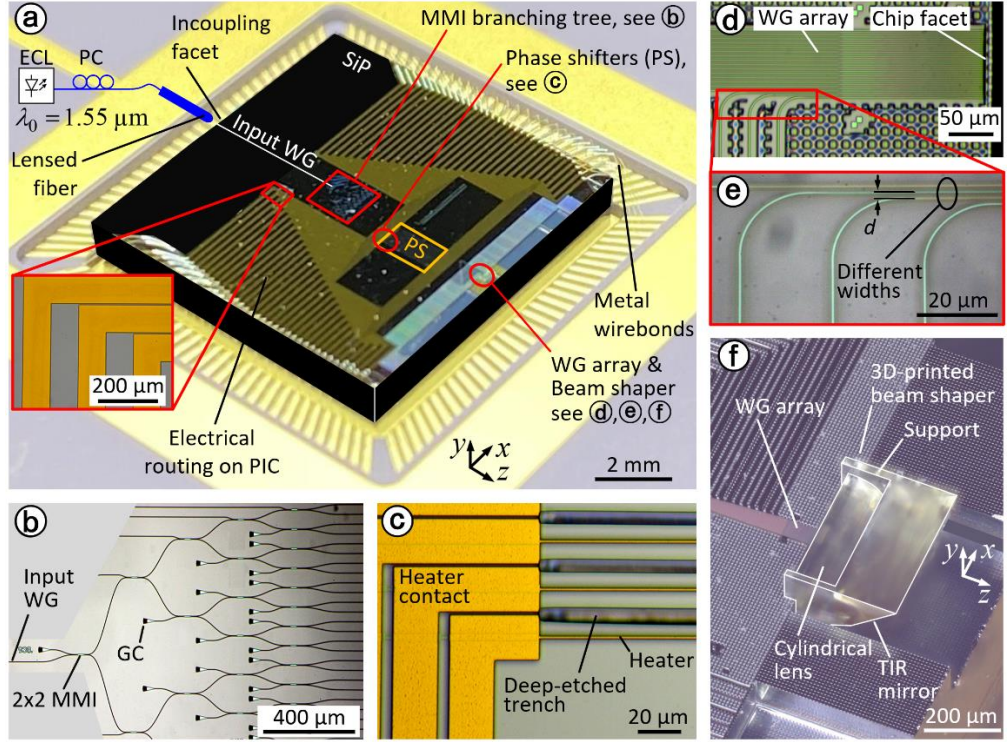

**Fig. 2**. Microscope images of the proof-of-concept assembly. **(a)** Photograph of the silicon photonic (SiP) chip in an electrical package. The photonic integrated circuit (PIC) contains an optical phased-array (OPA) with $N = 32$ channels having individual thermal phase shifters. These phase shifters are driven by DAC-controlled current sources via metal wirebonds. Laser light emitted by a benchtop-type external-cavity laser (ECL) is sent through a polarization controller (PC) and coupled to the input waveguide (WG) of the on-chip OPA using a lensed fiber. The light is then distributed to $N = 32$ phase shifters by a tree of cascaded MMI couplers. At the output, the light is emitted from tapered waveguide facets and sent through the 3D-printed beam shaper. The Inset shows a magnified part of the electrical wiring of the thermal phase-shifters. **(b)** Power-splitting tree, consisting of cascaded $2 \times 2$-multi-mode interference (MMI) couplers. Grating couplers (GC) provide auxiliary inputs, through which subsets of OPA channels can be addressed for testing purposes. **(c)** Thermal phase shifters, consisting of a resistive $1\,\mu m$-wide AlSiCu film on top of the silicon dioxide cladding. Each phase shifter section is 1.5 mm long, and the heater contacts are routed to bond pads (pitch $200\,\mu m$) along the PIC edges. Deep trenches etched down to the Si handle wafer prevent thermal crosstalk between neighboring heaters. **(d, e)** Densely packed OPA waveguides (pitch $d = 1.5\,\mu m$, tapered to a width $w = 350\,nm$ at the facet) lead to the PIC edge, ending inside the chip, approximately $1\,\mu m$ away from the facet. Subfigure **(e)** shows more details of these waveguide arrays. Adjacent waveguides have different widths and propagation constants for minimizing optical cross-coupling. **(f)** 3D-printed beam shaping element, including a total-internal reflection (TIR) mirror for out-of-plane emission, see Fig. 1. The beam shaper occupies an $x \times y \times z$ volume of $(380 \times 170 \times 300)\,\mu m^3$.

To avoid optical cross-talk between the parallel sections of the tightly spaced OPA output waveguides, two different waveguide widths $w_{WG}$ of 300 nm and 400 nm are alternated in this area, see Fig. 2(e). This leads to different propagation constants in neighboring waveguides, therefore minimizing optical coupling [16,18]. Towards the edge of the chip, the waveguides are first up-tapered to a common width of 480 nm, before being down-tapered to a final tip width of 350 nm, thereby guaranteeing identical far-field radiation characteristics (element factors, EF) of all OPA waveguides facets. The associated vertical and horizontal 3 dB beam divergences of a single emitter amount to $\Delta\Theta_{EF} = \pm 32°$ and $\Delta\Psi_{EF} = \pm 43°$, respectively, thereby covering the targeted $\pm 30°$ scanning range in $\Psi$-direction. To exclude cross-talk in the tapered waveguide sections as a relevant source of distortion in our devices, we assess the residual coupling between neighboring OPA waveguides both through simulations and through measurements. We find that the guided light remains confined to the respective waveguide without any leakage into neighboring ones. Note also that the OPA output waveguides end about 1 µm before the facet inside the chip, which needs to be taken into account for the lens design. We measure an approximately circular spot size with a 1/e²-diameter of the intensity distribution of $2r_{fac,meas} = 2.4\,\mu m$ at the facet of the chip, well in line with finite-difference time-domain (FDTD) simulations of the optical field emitted from the waveguide tip.

The overall structure of the 3D-printed beam shaper has been introduced in Fig. 1(b). As a key element of this structure, the shape of the cylindrical lens was designed and optimized using a home-made implementation of a wave-propagation algorithm based on a theory described in [19], that has previously been applied for design of 3D-printed freeform structures [20-22]. The optimized shape of the cylindrical lens shown in Fig. 2(f) is described by a $y$-dependent position $z$ of the lens surface, see also Fig. 1(b). This shape is parametrized by an even-order polynomial with four free coefficients $a_0$, $a_2$, $a_4$, $a_6$,

$$z(y) = a_0 + a_2 y^2 + a_4 y^4 + a_6 y^6 \qquad (1)$$

The $(y, z)$-origin of the coordinate system is at the center of the emission facet, see Fig. 1(b). After emitted by the lens, the beam passes through a prism-like redirection element that consists of a plane input facet pointing to the left, a plane TIR mirror surface at a 45° inclination, and a plane output facet pointing to the top. These plane facets only redirect the beam, but do not have any influence on the beam shape. For the optimization of the lens shape, we consider a two-dimensional field distribution in the plane $x = 0$ and assume that each of the OPA waveguide tips emits a Gaussian beam having a waist at the waveguide tip, where the 1/e²-diameter of the intensity distribution amounts to $2r_{tip} = 1.6\,\mu m$. Using the Gaussian mode field in the beam waist as an excitation for the wave propagation at $z = -d_{fac}$ and assuming a fixed value of $a_0 = 85\,\mu m$, we vary the coefficients $a_2$, $a_4$, and $a_6$ in Eq. (1) to maximize the overlap of the resulting emitted field with a two-dimensional Gaussian beam having a FWHM beam waist diameter of 25 µm, corresponding to a FWHM divergence angle of $\Delta\Theta_{FWHM} = 1.55°$. The position of the beam waist is adapted to yield maximum overlap with the emitted optical field. An optimum lens shape is obtained for $a_2 = -0.018\,\mu m^{-1}$, $a_4 = -2.60 \cdot 10^{-8}\,\mu m^{-3}$, $a_6 = -7.19 \cdot 10^{-10}\,\mu m^{-5}$. Note that the 1/e²-diameter mode-field diameter of $2r_{tip} = 1.6\,\mu m$ refers to the waist of the Gaussian beam right after the waveguide tip and is thus slightly smaller than the spot size of $2r_{fac,meas} = 2.4\,\mu m$, that was measured at the chip facet, i.e., $d_{fac} = 1\,\mu m$ away from the taper tip.

## 3. Device fabrication and characterization

The beam-shaping element is 3D-printed to the facet of the silicon photonic (SiP) chip *in-situ*, similar to the structures presented in [20,22,23]. We rely on an industry-grade two-photon lithography system (Sonata 1000, Vanguard Automation GmbH, Germany), which is also capable of printing photonic wirebonds (PWB) [12,13]. It is hence possible to fabricate package-level chip-chip and fiber-chip connections along with the 3D-printed beam shaper in a common, fully automated step of the underlying assembly process. A microscope image of the fabricated beam-shaping element is shown in Fig. 2(f). The structure consists of a photoresist (VanCore B, Vanguard Automation GmbH, refractive index $n = 1.53$ at $\lambda_0 = 1.55\,\mu\text{m}$) optimized for printing of micro-optical structures with high shape fidelity. In the 3D-printing process, the exact positioning of the cylindrical lens relative to the chip facet is crucial as it has a large influence on the collimation properties of the lens and on the direction of the collimated beam. Our machine achieves a placement accuracy of 30 nm such that positioning inaccuracies do not play a significant role.

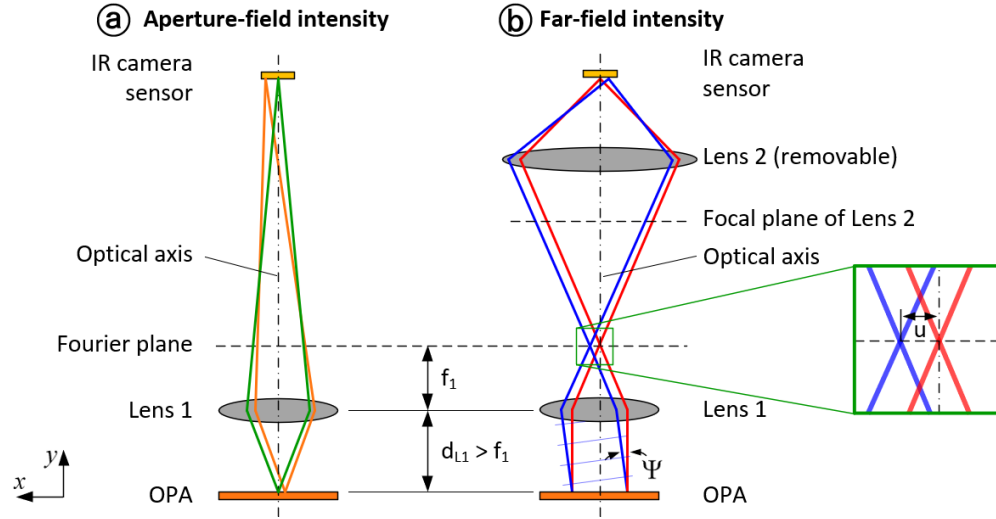

**Fig. 3.** Measurement setups for characterizing the intensity distributions in the emission aperture and in the far-field of the optical phased-array (OPA). **(a)** Measurement of the aperture-field intensity. The aperture field is imaged to an infrared (IR) camera sensor by Lens 1 having a focal length $f_1 = 15$ mm. The working distance is slightly larger than the focal length, $d_{L1} > f_1$, leading to a highly magnified image on the camera sensor. **(b)** Measurement of the far-field intensity. Each position $u$ in the Fourier plane of Lens 1 is associated with a specific angle $\Psi$ of a plane wave incident on Lens 1. Lens 2 is inserted between Lens 1 and the IR camera to image the intensity distribution in the Fourier plane of Lens 1 onto the camera sensor. Lens 2 also de-magnifies the field in the Fourier plane of Lens 1 to match the size of the sensor. For calibration of the setup, we use a single-mode fiber emitting a beam at an angle $\Psi$ (not shown). The fiber is rotated around its facet while recording the position $x$ of the center of the associated intensity distribution on the camera sensor.

Beside the cylindrical lens, the 3D-printed beam shaper comprises a TIR mirror that redirects the light to the top and that is held by support structures at each side, see Fig. 2(f).

After assembly of the system and 3D-printing of the beam-shaping element, the far-field intensity distribution of the light emitted by the optical phased-array (OPA) is characterized using the setup illustrated in Fig. 3, which similar to the one reported in [7]. Figure 3(a) shows the setup with a single lens between the OPA and the infrared (IR) camera sensor. The scanning angle $\Psi$ lies in the drawing plane of Fig. 3, whereas the out-of-plane direction corresponds to the angle $\Theta$, see Fig. 1(a) for the definition of the angles. Lens 1 is a microscope objective which maps the aperture field of the OPA to the camera sensor. The distance $d_{L1}$ between the lens and the OPA is slightly larger than the focal length $f_1$, $d_{L1} > f_1$, leading to a greatly magnified image of the aperture on the IR camera sensor. This setup helps in finding the emission aperture of the OPA and allows to investigate the intensity distribution in the aperture plane. For a subsequent measurement of the far-field intensity, an additional lens (Lens 2) is inserted, which images the intensity distribution in the back focal plane of Lens 1 to the camera sensor. Since each position $u$ in the back focal plane of Lens 1 is associated with a certain emission angle $\Psi$ with respect to the optical axis of the setup, the camera image directly reveals the angle-dependent intensity distribution of the OPA emission. Note that the field distribution in the back focal plane of Lens 1 would directly correspond to the Fourier transform of the aperture field if the OPA is precisely positioned in the front focal plane of Lens 1. However, since the distance $d_{L1}$ between the lens and the OPA slightly exceeds the focal length $f_1$ of Lens 1, only the modulus of the Fourier transform of the aperture field appears in the Fourier plane, but not its phase [24, Eq. 5-19]. Since we are only interested in the intensity distribution, this deviation does not play a role. In our setup, the intensity distribution in the back focal plane of Lens 1 can be captured within an area which has a length of $20\,\text{mm}$ along the horizontal direction ($\Psi$-direction) in Fig. 3 and a length of approximately $16\,\text{mm}$ in the direction perpendicular to it. This area is subsequently de-magnified to fit to the $9.5\,\text{mm} \times 7.6\,\text{mm}$ camera sensor. The relationship between the emission angles $\Psi$ and $\Theta$ and the corresponding position on the camera chip can be extracted by a calibration measurement in which the aperture of the OPA is replaced by the end-face of a single-mode fiber, which can be rotated in $\Psi$-direction while recording the position $x$ of the center of the associated intensity distribution on the camera sensor as a function of $\Psi$.

Due to manufacturing tolerances of the high-index-contrast silicon photonic waveguides, the optical phases of the fields emitted at the various facets of the OPA are subject to random variations. These variations come on top of the deterministically distinct phase shifts that the signals in the various channel experience on their ways through the 2x2 MMI splitters and the various waveguide sections with different design widths. To this end, we perform a one-time calibration process, in which we experimentally determine the phase shifter currents for every scan angle $\Psi$. This calibration procedure is similar to the one described in [8] and relies on maximizing the intensity at the targeted angle $\Psi$ by tuning all $N = 32$ phase shifters subsequently. The procedure is repeated for all steering angles, and the resulting heater currents are stored in a look-up table. In our experiments, we chose a steering angle increment of $\delta\Psi = 2°$. Figure 4(a) shows far-field intensity distributions of the calibrated OPA, measured for different steering angles $\Psi$ and then stacked on top of each other. The spot is steered across full range from $-30° \leq \Psi \leq 30°$, which is limited by the occurrence of secondary main lobes (grating lobes) due to higher diffraction orders, see Fig. 4(b).

Upon calibration, we characterize the divergence of the emitted beam. At small steering angles $|\Psi| \leq 10°$ the divergence in $\Psi$-direction, quantified via the full width at half maximum (FWHM) of the intensity distribution, is approximately $\Delta\Psi_{\text{FWHM,meas}} = 1.7°$, in good agreement with the theoretically expected value $\Delta\Psi_{\text{FWHM,theo}} = 1.64°$. The FWHM divergence in $\Theta$-direction at this steering angle is about the same, $\Delta\Theta_{\text{FWHM,meas}} = 1.7°$, in good agreement with the simulated imaging properties of the cylindrical lens resulting in $\Delta\Theta_{\text{FWHM,theo}} = 1.55°$. The far-field spot at $\Psi = 0°$ is thus approximately circular, and the beam quality factors are thus slightly larger than 1 both in $\Psi$- and $\Theta$-direction, $M_\Psi^2 = 1.7/1.64 = 1.04$ and $M_\Theta^2 = 1.7/1.55 = 1.1$. Insets (1) and (2) of Fig. 4(a) show exemplary far-field cuts in $\Theta$-direction (black dots) along with a Gaussian fit (red dashed line) at steering angles $\Psi = -20°$ and $\Psi = 10°$. Figure 4(b) shows the far-field cuts in $\Psi$-direction for the steering angles $\Psi = \pm 30°, \pm 16°, 0°$. At the edges of the steering range, i.e., for steering angles of $\Psi = \pm 30°$, we find secondary main lobes (grating lobes) at $\mp 32°$. These secondary main lobes occur at directions in which all emitted partial waves of the OPA interfere constructively in analogy to higher-order diffraction of conventional gratings. The directions are dictated by the waveguide pitch of $d = 1.5\,\mu m$, which is slightly smaller than a free-space wavelength $\lambda_0 = 1.55\,\mu m$. All other side lobes within the scan range $\Psi = -30°...30°$ are down by more than 10 dB compared to the main lobe across the entire scan range, see blue line in Fig. 4(b). The measured suppression ratio is slightly smaller than its theoretically predicted counterpart of approximately 13.2 dB. We attribute this deviation to residual roughness of the chip facet that still occurred in our devices due to non-optimum reactive-ion etching (RIE) processes for forming deep trenches along the dicing lanes of the wafer. Note that this roughness is not a fundamental problem, but a very specific issue of the fabrication processes used for our devices. Optimized etching processes can greatly reduce roughness and should hence lead to better radiation characteristics. These imperfections lead to different radiation characteristics, also called element factors (EF) of the various apertures, which, together with their spatial arrangement characterized by the so-called array factor (AF), form the total radiation pattern $EF \times AF$. Figure 4(c) depicts the FWHM divergences along the $\Psi$- and $\Theta$-direction as a function of steering angle $\Psi$. The FWHM divergence $\Delta\Theta$ in $\Theta$-direction is better than $1.9°$ for all steering angles and shows a slight systematic dependence on the steering angle $\Psi$. This is caused by the fact that the effective focal length of the cylindrical lens changes when then lens is traversed at a steering angle $\Psi \neq 0$. The rather random variations of the FWHM divergence $\Delta\Psi$ in $\Psi$-direction is attributed to the aforementioned roughness of the chip facet.

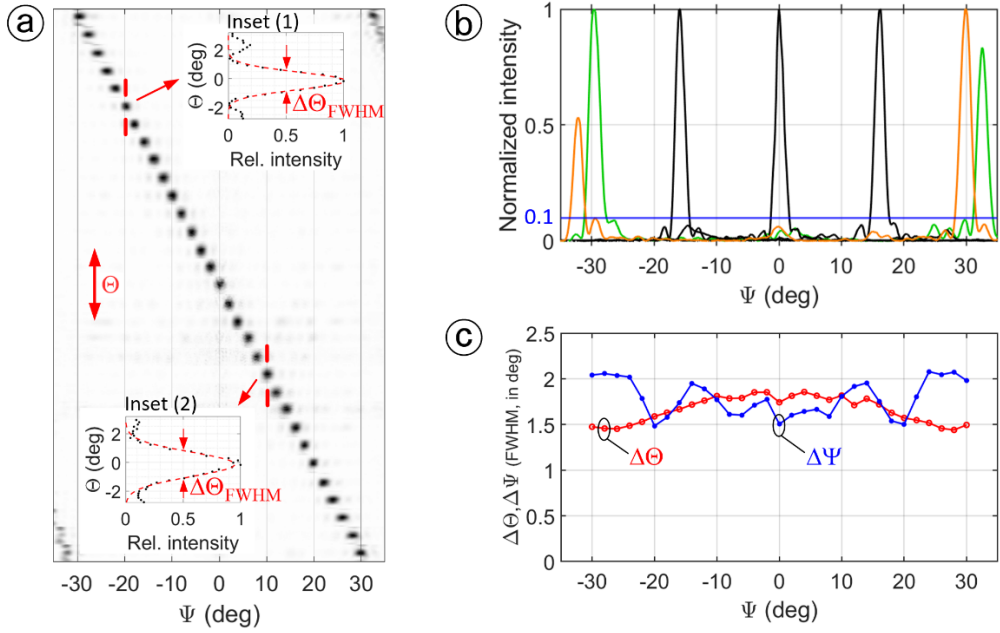

**Fig 4**. Far-field intensity measurement of the optical phased-array (OPA) with 3D-printed beam shaper in a setup according to Fig. 3(b). The elevation angle $\Theta$ and the steering angle $\Psi$ are indicated in Fig. 1. **(a)** Measured far-field intensities for different steering angles $\Psi$ and $\Theta = 0°$, stacked on top of each other. The beam is scanned across the grating-lobe-free steering range of $\Psi_{max} = \pm 30°$ in steps of $\delta\Psi = 2°$. Exemplary far-field cuts in $\Theta$-direction for steering angles $\Psi = -20°$ and $\Psi = 10°$ are shown in Insets 1 and 2. Black dots indicate the measured data points, and the red dashed line shows a Gaussian fit. The full width-at-half-maximum (FWHM) $\Delta\Theta_{FWHM}$ is marked by arrows. **(b)** Exemplary far-field intensity distributions along the steering direction ($\Psi$-direction). When steered to $\Psi = \pm 30°$ secondary main lobes appear at $\Psi = \mp 32°$, in analogy to higher diffraction orders of conventional gratings – the associated intensity distributions are depicted in green and orange for better visibility. The positions of these secondary main lobes are dictated by the waveguide pitch $d = 1.5\,\mu m$, which is only slightly smaller than the vacuum wavelengths ($\lambda_0 = 1.55\,\mu m$). All other side lobes are down by at least 10 dB compared to the main lobe, see horizontal blue line. **(c)** FWHM divergence in $\Theta$-direction and $\Psi$-direction $\Delta\Psi_{FWHM}$ for all steering angles. The $\Theta$-divergence $\Delta\Theta_{FWHM}$ is smaller than $1.9°$ for all steering angles $\Psi$ and depends slightly on $\Psi$ since the employed cylindrical lens is traversed a lateral angle for $\Psi \neq 0$.

We also estimated the emission efficiency of the 3D-printed beam shaper and the associated on-chip interface. To this end, we couple light into the input waveguide of our OPA using a lensed fiber, see Fig. 2(a). We steer the OPA to an emission direction of $\Psi = 0°$ and use an integrating sphere (Thorlabs S145C) with a round input aperture to measure the optical power emitted by the 3D-printed beam shaper into a circular cone with an opening half-angle of 15° (full angle 30°), centered about the beam axis. Stray light from the chip is shielded by an appropriate cover. From the lensed fiber to the integrating sphere, we measure, after a calibration step, a power loss of 20.8 dB. Coupling loss from the lensed fiber to the chip amounts to 7.3 dB, and approximately 3.3 dB of loss are added by an MMI that is included into the on-chip input waveguide for test purposes. This MMI couples 50% of the incoming light to a reference port and features an excess insertion loss of 0.27 dB. Similarly, the subsequent five stages of the MMI-based 1:32 splitter tree contribute excess losses of approximately 0.27 dB per stage, which leads to a total loss of 1.35 dB, obtained by relating the power at the input of the splitter tree to total power at all its outputs. The average waveguide attenuation amounts to 0.43 dB/mm, which leads to a total loss of approximately 4.7 dB along each of the 11 mm-long signal paths. Reflection losses at the waveguide tips and emission into unwanted secondary main lobes (grating lobes) add another loss of approximately 2.7 dB. Disregarding the unknown roughness loss of the radiating facets, this leads to an OPA loss of approximately 19.4 dB, leaving approximately 1.4 dB of loss (72% emission efficiency) for the current implementation of the 3D-printed beam shaper. The Fresnel reflections at the three interfaces of the 3D-printed beam shaper including the prism-like redirection element amount to approximately 0.6 dB. We attribute the remaining losses of approximately 0.8 dB to fabrication inaccuracies of the 3D-printed elements, to the aforementioned roughness of the chip facet, and to scattering of light from the waveguide tips to directions outside the capture range of the 3D-printed beam shaper. Still, the 1.4 dB of loss measured for the current structures is significantly lower than the 5 dB of loss that were measured for single-layer grating couplers [25] and can well compete with the losses of 1.33 dB (0.97 dB directionality + 0.36 dB loss due to incomplete radiation of power) and 2.06 dB (1.04 dB directionality + 1.02 dB loss due to incomplete radiation of power) that were obtained for best-in-class $Si_3N_4$-based dual-level fishbone and chain gratings [11]. In contrast to such gratings, our 3D-printed beam-shaping elements neither require excessive on-chip area nor do they rely on complex platform-specific sequences of patterning and deposition steps. Based on the sub-1 dB loss demonstrated for beam shapers printed to edge-emitting InGaAsP lasers [23], we believe that further optimization of the emission interface, using, e.g., sub-wavelength grating (SWG) waveguides [26-28], in combination with smooth chip facets and improved fabrication processes for the 3D-printed beam shaper will eventually permit emission losses below 1 dB.

## 4. Summary and outlook

We demonstrate 3D-printed facet-attached beam shapers as an attractive approach to beam forming and out-of-plane deflection of light in integrated optical phased-arrays (OPA). The elements are *in-situ* printed to the chip facet using high-resolution multi-photon lithography, making the concept applicable to any integration platform that offers edge-emitting facets. We demonstrate the viability of the scheme in a proof-of-concept experiment using an edge-emitting silicon photonic OPA. Within the steering range of $\pm 30°$, we achieve a full-width-half-maximum beam divergence of less than 2.1° along the steering direction and of 1.9° in the direction perpendicular to it. The steering range is limited by the 1.5 µm pitch of the edge-emitting waveguide facets and can be further increased by using smaller pitches. Our current device generation exhibits emission efficiencies of 72 % (1.4 dB loss) and can well compete with best-in-class $Si_3N_4$-based dual-level gratings [11], which occupy considerable on-chip footprint and which rely on rather complex sequences of platform-specific patterning and deposition steps. We believe that losses below 1 dB will become feasibly with further optimization of designs and fabrication processes.

**Funding**

Bundesministerium für Bildung und Forschung (BMBF), projects "OPALID" (#13N14589) and "DiFeMiS" (#16ES0948); Deutsche Forschungsgemeinschaft (DFG), Excellence Cluster "3D Matter Made to Order" (3DMM2O, EXC-2082/1 – 390761711) and Collaborative Research Center "WavePhenomena" (CRC 1173); European Research Council, ERC Consolidator Grant TeraSHAPE (#773248); Alfried Krupp von Bohlen und Halbach Foundation.

**Disclosures**

P.-I. D. and C. K. are co-founders and shareholders of Vanguard Automation GmbH (Karlsruhe, Germany) and Vanguard Photonics GmbH (Karlsruhe, Germany), start-up companies focusing on commercializing 3D nano-printing techniques in the field of integrated optics. S. T. S. and P.-I. D. are employees of Vanguard Automation GmbH.

**Data availability**

Data underlying the results presented in this paper are not publicly available at this time but may be obtained from the authors upon reasonable request.

**References**

1. C. V. Poulton, A. Yaacobi, D. B. Cole, M. J. Byrd, M. Raval, D. Vermeulen, and M. R. Watts, "Coherent solid-state LIDAR with silicon photonic optical phased arrays," Opt. Lett. **42**(20), 4091-4094 (2017).
2. L. Zhang, Y. Li, B. Chen, Y. Wang, H. Li, Y. Hou, M. Tao, Y. Li, Z. Zhi, X. Liu, X. Li, Q. Na, Q. Xie, M. Zhang, X. Li, F. Gao, X. Luo, G.-Q. Lo, and J. Song, "Two-dimensional multi-layered SiN-on-SOI optical phased array with wide-scanning and long-distance ranging," Opt. Express **30**(4), 5008-5018 (2022).
3. C.-P. Hsu, B. Li, B. Solano-Rivas, A. R. Gohil, P. H. Chan, A. D. Moore, and V. Donzella, "A Review and Perspective on Optical Phased Array for Automotive LiDAR," IEEE J. Sel. Top. Quantum Electron. **27**(1), 8300416 (2021).
4. C. V. Poulton, M. J. Byrd, E. Timurdogan, P. Russo, D. Vermeulen, and M. R. Watts, "Optical Phased Arrays for Integrated Beam Steering," in Proceedings of IEEE 15th International Conference on Group IV Photonics (IEEE 2018), pp.1-2.
5. M. Raval, C. V. Poulton, and M. R. Watts, "Unidirectional waveguide grating antennas with uniform emission for optical phased arrays," Opt. Lett. **42**(13), 2563-2566 (2017).
6. Y. Luo, Z. Nong, S. Gao, H. Huang, Y. Zhu, L. Liu, L. Zhou, J. Xu, L. Liu, S. Yu, and X. Cai, "Low-loss two-dimensional silicon photonic grating coupler with a backside metal mirror," Opt. Lett. **43**(3), 474-477 (2018).
7. J. Sun, E. Timurdogan, A. Yaacobi, E. S. Hosseini, and M. R. Watts, "Large-scale nanophotonic phased array," Nature **493**, 195-199 (2013).
8. D. Kwong, A. Hosseini, J. Covey, Y. Zhang, X. Xu, H. Subbaraman, and R. T. Chen, "On-chip silicon optical phased array for two-dimensional beam steering," Opt. Lett. **39**(4), 941-944 (2014).
9. T. Aalto, M. Cherchi, M. Harjanne, S. Bhat, P. Heimala, F. Sun, M. Kapulainen, T. Hassinen, and T. Vehmas, "Open-Access 3-μm SOI Waveguide Platform for Dense Photonic Integrated Circuits," IEEE J. Sel. Top. Quantum Electron **2**(5), 8201109 (2019).
10. S.-M. Kim, E.-S. Lee, K.-W. Chun, J. Jin, and M.-C. Oh, "Compact solid-state optical phased array beam scanners based on polymeric photonic integrated circuits," Sci. Rep. **11,** 10576 (2021).
11. B. Chen, Y. Li, L. Zhang, Y. Li, X. Liu, M. Tao, Y. Hou, H. Tang, Z. Zhi, F. Gao, X. Luo, G. Lo, and J. Song, "Unidirectional large-scale waveguide grating with uniform radiation for optical phased array," Opt. Express **29**(13), 20995-21010 (2021).

12. M. R. Billah, M. Blaicher, T. Hoose, P.-I. Dietrich, P. Marin-Palomo, N. Lindenmann, A. Nesic, A. Hofmann, U. Troppenz, M. Moehrle, S. Randel, W. Freude, and C. Koos, "Hybrid integration of silicon photonic circuits and InP lasers by photonic wire bonding," Optica **5**(7), 876-883 (2018).
13. M. Blaicher, M. R. Billah, J. Kemal, T. Hoose, P. Marin-Palomo, A. Hofmann, Y. Kutuvantavida, C. Kieninger, P.-I. Dietrich, M. Lauermann, S. Wolf, U. Troppenz, M. Moehrle, F. Merget, S. Skacel, J. Witzens, S. Randel, W. Freude, and C. Koos, "Hybrid multi-chip assembly of optical communication engines by in situ 3D nano-lithography," Light Sci. Appl. **9**, 71 (2020).
14. IMS CHIPS, institute website https://www.ims-chips.com/ (last downloaded June 21st, 2022).
15. M. I. Skolnik, *Introduction to Radar systems* (McGraw-Hill, 2003).
16. C. T. Phare, M. C. Shin, S. A. Miller, B. Stern, and M. Lipson, "Silicon Optical Phased Array with Grating Lobe-Free Beam Formation Over 180 Degree Field of View," in Conference on Lasers and Electro-Optics, OSA Technical Digest (online) (Optica Publishing Group, 2018), paper SM3I.2.
17. M. R. Kossey, C. Rizk, and A. C. Foster, "End-fire silicon optical phased array with half-wavelength spacing," APL Phot. **3**, 011301 (2018).
18. L. Wang, Z. Chen, H. Wang, A. Liu, P. Wang, T. Lin, X. Liu, and H. Lv, "Design of a low-crosstalk half-wavelength pitch nano-structured silicon waveguide array," Opt. Lett. **44**(13), 3266-3269 (2019).
19. S. Schmidt, T. Tiess, S. Schröter, R. Hambach, M. Jäger, H. Bartelt, A. Tünnermann, and H. Gross, "Wave-optical modeling beyond the thin-element approximation," Opt. Express **24**(26), 30188-30200 (2016).
20. Y. Xu, A. Kuzmin, E. Knehr, M. Blaicher, K. Ilin, P.-I. Dietrich, W. Freude, M. Siegel, and C. Koos, "Superconducting nanowire single-photon detector with 3D-printed free-form microlenses," Opt. Express **29**(17), 27708-27731 (2021).
21. M. Trappen, M. Blaicher, P.-I. Dietrich, C. Dankwart, Y. Xu, T. Hoose, M. R. Billah, A. Abbasi, R. Baets, U. Troppenz, M. Theurer, K. Wörhoff, M. Seyfried, W. Freude, and C. Koos, "3D-printed optical probes for wafer-level testing of photonic integrated circuits," Opt. Express **28**(25), 37996-38007 (2020).
22. P. Maier, Y. Xu, M. Lauermann, A. Henniger-Ludwig, H. Kapim, M. Trappen, T. Kind, A. Weber, M. Blaicher, P.-I. Dietrich, C. Wurster, S. Randel, W. Freude, and C. Koos, "3D-Printed Optical Elements for Coupling of VCSEL and Photodiode Arrays to Multi-Core Fibers in an SFP Transceiver Assembly," published in 2022 Optical Fiber Communication Conference and Exhibition (OFC), paper W2A.1.
23. P.-I. Dietrich, M. Blaicher, I. Reuter, M. Billah, T. Hoose, A. Hofmann, C. Caer, R. Dangel, B. Offrein, U. Troppenz, M. Moehrle, W. Freude, and C. Koos, "In situ 3D nanoprinting of free-form coupling elements for hybrid photonic integration," Nat. Photonics **12**, 241-247 (2018).
24. J. W. Goodman, *Introduction to Fourier Optics* (Roberts & Company Publishers, 2005).
25. C. V. Poulton, M. J. Byrd, M. Raval, Z. Su, N. Li, E. Timurdogan, D. Coolbaugh, D. Vermeulen, and M. R. Watts, "Large-scale silicon nitride nanophotonic phased arrays at infrared and visible wavelengths," Opt. Lett. **42**(1), 21-24 (2017).
26. P. Cheben, J. H. Schmid, S.Wang, D.-X. Xu, M. Vachon, S. Janz, J. Lapointe, Y. Painchaud, and M.-J. Picard, "Broadband polarization independent nanophotonic coupler for silicon waveguides with ultra-high efficiency," Opt. Express **23**(17), 22553-22563 (2015).
27. Y. Xiao, Y. Xu, Y. Dong, B. Zhang, and Y. Ni, "A 60 μm-Long Fiber-to-Chip Edge Coupler Assisted by Subwavelength Grating Structure with Ultralow Loss and Large Bandwidth," Photonics 2022 **9**, 413 (2022).
28. R. Halir, A. Ortega-Monux, D. Benedikovic, G. Z. Mashanovich, J. G. Wanguemert-Perez, J. H. Schmid, I. Molina-Fernandez, and P. Cheben, "Subwavelength-Grating Metamaterial Structures for Silicon Photonic Devices," in Proceedings of the IEEE **106** (12), 2144-2157 (2018).